%% file: main.tex
\documentclass[runningheads]{llncs}
\pdfoutput=1
\usepackage[T1]{fontenc}
\PassOptionsToPackage{hyphens}{url}
\usepackage{graphicx}
\usepackage{listings}
\usepackage{multirow}
\usepackage{booktabs}
\usepackage{array}
\usepackage[
  hidelinks,
  pdfauthor={Ralf Gundelach, Michael Mühlhauser, Dominik Herrmann},
  pdftitle={Detecting Bot Detection: Prevalence, Techniques, and Implications for Web Measurement Research}
]{hyperref}
\usepackage{xurl}
\begin{document}
\title{Detecting Bot Detection: Prevalence, Techniques, and Implications for Web Measurement Research}
\titlerunning{Detecting Bot Detection}

\author{Ralf Gundelach \and
Michael Mühlhauser \and
Dominik Herrmann}

\authorrunning{R. Gundelach et al.}

\institute{University of Bamberg\\
\email{\{rg.psi,mm.psi,dh.psi\}@uni-bamberg.de}}

\maketitle%
\begin{abstract}
Browser automation frameworks are essential tools for security and privacy research on the web, yet bot detection scripts increasingly probe their artifacts, threatening measurement validity as automated browsers may be blocked or served different content.
Prior work measures detection deployment, while we measure blocking-induced sample loss. 
Through a literature survey of top-tier security, privacy, and web measurement venues, we find that 83\% of papers omit any discussion of bot detection blocking.
To address this gap, we conduct a measurement study of 10{,}000 websites across four browser configurations (40K page visits in total) to quantify detection prevalence and employed techniques.
Using custom instrumentation to detect when sites probe for automation, we develop a taxonomy of bot detection techniques and measure how often they appear in practice.
Chromium headless encounters a 15\% soft block rate compared to 7\% for other configurations.
Across all conditions, 82\% of blocks are attributable to bot detection (59\% vendor-confirmed, 23\% inferred from condition-dependent blocking), predominantly by providers with integrated bot detection such as Cloudflare (37\% block rate) and Akamai (26\%).
A header spoofing experiment establishes that 75\% of Chromium-headless-only blocks are caused by header-level signals alone, yet JavaScript-based environment probing is more extensive than current blocking rates suggest.
These findings demonstrate that bot detection creates systematic, provider-correlated sample loss that the web measurement community neither measures nor reports.
The downstream effect on specific measurement outcomes remains future work.
\keywords{Bot Detection \and Browser Fingerprinting \and Browser Automation \and Web Crawling \and Reproducibility}
\end{abstract}

\input{sections/introduction.tex}

\input{sections/bg-and-related-work.tex}

\input{sections/literature-survey.tex}

\input{sections/dns-census.tex}

\input{sections/bot-detection-measurement.tex}

\section{Discussion and Implications}\label{sec:discussion}

\paragraph{Why CDN-correlated bias differs from other sampling biases.}
Unlike rank-based or geographic biases, which can be addressed through stratified sampling, CDN-correlated attrition is invisible to the sampling process: researchers cannot determine which CDN a target site uses before crawling, and therefore cannot compensate for differential blocking rates.
Because providers with default-on bot detection block at 26--37\% while others range from 0--16\% (Table~\ref{tab:as_domain_count}), the resulting sample loss correlates with infrastructure features such as security posture and bot management defaults, not with the content properties that measurement studies aim to capture.

\paragraph{Implications for published results.}
When a crawler receives a challenge page or HTTP~403, the tracking scripts, third-party requests, and advertising infrastructure on the actual page go unobserved.
To illustrate: Iqbal et~al.~\cite{DBLP:conf/sp/IqbalES21} report fingerprinting on over a quarter of the Alexa Top~10K. 
Our data shows that even Firefox-based crawlers are blocked on 7\% of sites, predominantly by Cloudflare, which itself deploys fingerprinting scripts on its challenge pages.
Whether blocked sites exhibit higher or lower fingerprinting prevalence cannot be determined without a human-browsing baseline, but the bias is systematic and CDN-correlated rather than random.
A replication of such a study today would systematically miss all Cloudflare-blocked sites, potentially undercounting fingerprinting on sites that deploy both bot detection and tracking, or overcounting it if challenge pages are mistakenly analyzed as regular content.

\paragraph{Detection infrastructure exceeds current blocking.}
The detection infrastructure already deployed is more extensive than blocking rates suggest: \texttt{navigator.\allowbreak{}webdriver} is probed on 34\% of sites, and Tier~1 automation honeypots are checked on 46\%.
The AI scraping boom accelerates this trend: agentic AI traffic grew 6{,}900\% year-over-year in 2025\footnote{HUMAN Security, ``Agentic Commerce Traffic,'' \url{https://www.humansecurity.com/learn/blog/agentic-commerce-traffic-black-friday-cyber-monday/}}, prompting CDN vendors to deploy default-on blocking of AI crawlers\footnote{Namecheap, ``Could Cloudflare's Bot Blocker Impact Your Brand's Discoverability?'' \url{https://www.namecheap.com/blog/could-cloudflares-bot-blocker-impact-your-brands-discoverability/}} and introduce pay-per-crawl monetization models\footnote{Cloudflare, ``Introducing Pay Per Crawl,'' \url{https://blog.cloudflare.com/introducing-pay-per-crawl/}}.
Academic web measurement relies on the same automation toolchain (Playwright, Puppeteer, OpenWPM), meaning that researchers face increasingly aggressive blocking as a side effect of these vendor countermeasures.

\paragraph{Limitations.}
Beyond the instrumentation caveats discussed in Section~\ref{sec:methods}, our study has several scope limitations.
All scans originate from a single vantage point (a European datacenter) at a single point in time; multi-vantage-point measurements would reveal whether blocking varies by geography or IP reputation.
Our sample is limited to the Tranco Top~10K; we visit only homepages.
Our methodology captures only explicit HTTP error responses (403/429/503); content degradation within HTTP~200 responses is outside our observational scope.
Finally, we lack a true human-browsing baseline; our headed conditions approximate but do not fully replicate manual browsing.
Despite these limitations, our core findings hold: the paired $2 \times 2$ design controls for site-level and temporal variation, the header spoofing experiment establishes causality for the dominant blocking mechanism, and the CDN-correlation finding is independent of vantage point or sample size.

\section{Conclusion}\label{sec:conclusion}
Our measurement of 10{,}000 websites across four browser configurations shows that bot detection creates systematic sample loss: soft block rates range from 7--15\% depending on configuration, with 82\% attributable to bot detection.
Because blocking is provider-driven, not rank-driven, the resulting attrition is CDN-correlated and cannot be addressed through stratified sampling, yet 83\% of measurement papers we surveyed do not report it.
As header-level signals are patched and AI-driven scraping intensifies vendor countermeasures, blocking rates will likely increase, making this an urgent methodological concern.

We recommend that future measurement studies at minimum:
(1)~report the browser engine, version, and display mode used;
(2)~quantify the blocking or failure rate observed during crawling;
(3)~document the CDN composition of their sample, as infrastructure provider is a confound for blocking;
(4)~describe any stealth measures employed or deliberately omitted.
Longitudinal replication and a human-browsing baseline to determine the direction of the resulting bias are natural next steps.

\begin{credits}
\subsubsection{\ackname} IP address data is powered by IPinfo (\url{https://ipinfo.io}). 
We used Claude (Anthropic) during the development of our scanning infrastructure for debugging, code review, and drafting SQL analysis queries. 
All technical decisions, experimental design, and interpretation of results were made by the authors. The final manuscript was proofread with LLM assistance.

\subsubsection{\discintname} The authors have no competing interests to declare that are relevant to the content of this article.
\end{credits}
\bibliographystyle{splncs04}
\bibliography{literature}
\end{document}

%% file: sections/introduction.tex
\section{Introduction}\label{sec:introduction}

Browser automation frameworks have been crucial for web security and privacy research. 
They enable large-scale measurements of tracking~\cite{DBLP:conf/ccs/EnglehardtN16} and vulnerability scanning~\cite{DBLP:conf/acsac/KleinMBKJ22}. 
However, the validity of these measurements is endangered by bot detection, a factor whose implications for measurement validity are not yet well characterized. 

Website providers deploy anti-bot defenses for legitimate purposes, such as credential stuffing mitigation and scraping protection~\cite{DBLP:conf/dimva/AzadSLN20}.
These defenses, either in the form of community-driven projects or commercial anti-bot services offered by companies like Cloudflare, Akamai, and PerimeterX/HUMAN\footnote{Cloudflare, ``Bot Solutions,'' \url{https://developers.cloudflare.com/bots/}; Akamai, ``Bot Manager,'' \url{https://akamai.com/products/bot-manager}; HUMAN Security, ``Bot Protection,'' \url{https://humansecurity.com/platform/solutions/bot-detection-mitigation}}, have become ubiquitous across the web.
This dynamic is compounded by the rise of large language models (LLMs): bot detection vendors now explicitly market capabilities to block AI scrapers\footnote{Cloudflare, ``Declare your AIndependence,'' \url{https://blog.cloudflare.com/declaring-your-aindependence-block-ai-bots-scrapers-and-crawlers-with-a-single-click/}; HUMAN Security, ``Controlling AI-driven Content Scraping,'' \url{https://www.humansecurity.com/learn/blog/controlling-ai-driven-content-scraping-with-human/}}, tightening defenses that also affect researchers using the same tools.
The deployment of bot detection measures creates a fundamental threat to measurement validity:
automated browsers may be blocked, rate-limited, or served different content than human users, causing previous measurement studies to silently reflect only the subset of sites that did \emph{not} block automation.

Prior work has examined bot detection from specific angles, like reverse engineering individual detectors~\cite{DBLP:conf/esorics/JonkerKV19}, measuring framework detectability~\cite{DBLP:conf/conext/KrumnowJK22}, and analyzing cloaking in adversarial settings~\cite{DBLP:journals/ieeesp/ZhangOCSJWSKBWS22}. 
However, no study has systematically measured how bot detection affects web measurement research at scale with respect to blocking outcomes. 

We address this gap through three contributions:

\begin{enumerate}
\item We establish a measurement blind spot through a literature survey of 81 papers at top-tier venues, finding that 83\% omit any discussion of bot detection blocking.
\item We map the domains of the Tranco Top~1M~\cite{DBLP:conf/ndss/PochatGTKJ19}, a toplist used in many measurement studies, to their autonomous systems. 
We show that a majority of websites are served through CDNs with bot detection capabilities, unavoidably exposing crawling experiments to detection infrastructure. 
\item We scan the top 10{,}000 sites of the Tranco Top~1M across four browser conditions and find that blocking is provider-driven: 
Cloudflare and Akamai block at the highest rates (37\% and 26\%), while providers without default-on bot detection range from 0--16\%.
A header spoofing experiment (Section~\ref{sec:rq1}) shows that 75\% of Chromium-headless blocks are caused by header-level signals alone, yet browser environment probing is more extensive than current blocking rates suggest, with 46\% of sites checking JavaScript properties that are only present in automated browsers.
\end{enumerate}

Our findings demonstrate that bot detection creates systematic sample loss correlated with CDN infrastructure that the community neither measures nor reports, with direct consequences for the validity of published results.
Our crawler, dataset, analysis notebooks, and per-paper literature survey codings are publicly available.\footnote{\url{https://doi.org/10.5281/zenodo.20334022}} Per-site screenshots are excluded from the released dataset for copyright reasons.

The rest of this paper is structured as follows: 
Section~\ref{sec:related-work} discusses background and related work. 
Section~\ref{sec:literature-survey} presents our literature survey. 
Section~\ref{sec:dns-census} quantifies CDN prevalence through DNS resolution of the Tranco Top~1M. 
Section~\ref{sec:measurement} describes our measurement methodology and results. 
Section~\ref{sec:discussion} discusses implications, recommendations, and limitations. 
Section~\ref{sec:conclusion} concludes this work.

%% file: sections/bg-and-related-work.tex
\section{Background \& Related Work}\label{sec:related-work}

Research on detecting web measurement frameworks is limited.
Demir et al.~\cite{DBLP:conf/www/DemirGUWHP22} surveyed 117 measurement studies and found that undocumented bot evasion strategies and varying crawler setups severely impact results.
We draw on related work in browser fingerprinting measurement, which shares our methodology of intercepting API accesses at scale, and on web bot detection research, where prior studies have identified detection scripts through string matching, examined the fingerprinting surface of OpenWPM, and evaluated anti-bot services from a defender's perspective.

\subsection{Browser Fingerprinting}

Englehardt and Narayanan developed OpenWPM, an open-source web measurement framework, which they used to conduct a large-scale measurement of tracking on 1M websites~\cite{DBLP:conf/ccs/EnglehardtN16}. 
By instrumenting JavaScript APIs (Canvas, WebRTC, AudioContext, and Battery API), they were able to measure fingerprinting by third parties. 
Iqbal et al.~\cite{DBLP:conf/sp/IqbalES21} extended OpenWPM to track access to the Performance API and WebGL, as they found that these APIs were also used by popular fingerprinting scripts. 
They applied a syntactic-semantic based approach and found that browser fingerprinting is present on over a quarter of the Alexa Top 10K and over 10\% of the Top 100K. 
Vastel et al.~\cite{DBLP:conf/uss/VastelLRR18} introduced FP-Scanner, a fingerprint scanner that can detect inconsistencies in browser fingerprints and whether the presented fingerprint is genuine or not. 
Bahrami et al.\ employed a graph-based approach on historical snapshots of the Alexa Top 100K to discover previously unknown fingerprinting APIs through clustering~\cite{DBLP:journals/popets/BahramiIS22}.
Boussaha et al.\ used dynamic taint tracking to measure fingerprinting severity across the Tranco Top 100K, finding that 8\% of domains perform high-severity fingerprinting~\cite{DBLP:journals/popets/BoussahaHBRRKJCAB24}. 
However, these automated measurements likely underestimate the prevalence of fingerprinting. 
Annamalai et al.~\cite{DBLP:conf/www/AnnamalaiCB25} demonstrated that automated crawls miss up to 45\% of the fingerprinting websites encountered by real users, largely due to bot detection mechanisms and consent banners blocking crawler access. 

\subsection{Web Bot Detection}

Jonker et al. reverse-engineered a commercial client-side web bot detector~\cite{DBLP:conf/esorics/JonkerKV19}. 
They identified behavior-based detection using event handlers, code injection routines for script-website communication, and detection based on DOM properties. 
By extending the open-source library fingerprint2.js, they determined the fingerprinting surface of 14 web bots. 
Using derived keyword patterns, they scanned the Alexa Top 1M and found 127{,}799 sites with matching scripts. Detection prevalence was higher in the top 100K (15.7K sites) compared to lower ranks (averaging 12.7K per 100K). 
Jueckstock et al. developed VisibleV8, which instruments the V8 JavaScript engine at the browser level to log all native API accesses~\cite{DBLP:conf/imc/JueckstockK19}. 
During a crawl of the Alexa Top~50K, they discovered that 29\% of websites loaded scripts that were probing for 46 JavaScript namespace artifacts associated with automated browser platforms. 
Krumnow et al. analyzed the detectability of OpenWPM, the resilience of its data recording, and the prevalence of OpenWPM detection~\cite{DBLP:conf/conext/KrumnowJK22}. They scanned the Tranco Top 100K and determined that 14\% of the front pages detected OpenWPM, rising to 19\% when sub-pages were visited.
Vastel et al.~\cite{vastelFPCrawlersStudyingResilience2020} provided the first systematic analysis of fingerprinting-based crawler detection: they crawled the Alexa Top 10K with Chromium headless and identified 291~sites (2.9\%) that block crawlers, of which 93 use browser fingerprinting, and further evaluated evasion resilience by incrementally spoofing crawler attributes.

Other work takes an adversarial perspective.
Azad et al.~\cite{DBLP:conf/dimva/AzadSLN20} evaluated commercial anti-bot services and found that while basic bots (Python, PhantomJS) are blocked on over 75\% of protected sites, less common browsers (Safari, mobile Chrome) bypass up to 82\%.
Venugopalan et al.~\cite{DBLP:conf/imc/VenugopalanMAWK25} solicited traffic from 20~bot services against a honey site and found evasion rates of 45--53\%; they proposed FP-Inconsistent, which detects bots through fingerprint inconsistencies, reducing evasion by 45--48\%.

Prior work has established that bot detection scripts are widely deployed~\cite{DBLP:conf/esorics/JonkerKV19,DBLP:conf/imc/JueckstockK19,DBLP:conf/conext/KrumnowJK22}, but these studies measure detection \emph{presence}, e.\,g., whether a site loads with capabilities for detecting automated browsers, not detection \emph{impact}:
whether the site actually blocks automated browsers, serves altered content, or behaves differently across browser configurations as a consequence of the detection.
The distinction matters, as deploying a detection script does not imply blocking: Krumnow et al. found that many sites probe for automation signals without necessarily acting on the result. 
However, when active mitigation is encountered, the impact on measurement validity can be significant. 
Jueckstock et al.~\cite{DBLP:conf/www/JueckstockSSBPV21} found that crawler configurations and anti-bot defenses introduce significant measurement bias, altering the advertising and tracking domains encountered by up to 19\%. 
Vastel et al.\ take a step further by measuring actual blocking, but their resilience evaluation covers 40 selected sites with a single browser engine.
Our study shifts the focus from evasion resilience to infrastructure-level prevalence: we measure blocking outcomes across 10{,}000 sites and four browser conditions, identifying which providers drive blocking and establishing through a causal experiment which signals cause it.

%% file: sections/literature-survey.tex
\section{Bot Detection Reporting in Web Measurement Studies}\label{sec:literature-survey}

This section addresses our first contribution by establishing a measurement blind spot in web measurement research.
To motivate the need for better bot detection transparency, we conducted a systematic literature review of bot detection reporting practices in recent crawling studies.
We describe our methodology in Section~\ref{sec:litsurvey:methods} and present the results in Section~\ref{sec:litsurvey:results}.

\subsection{Methods}\label{sec:litsurvey:methods}

We surveyed papers published between 2020 and 2025 at top-tier security, privacy, and web measurement venues: USENIX Security, IEEE S\&P, CCS, NDSS, WWW, IMC, and PETS. 
For each paper, we read the title and abstract to determine whether it involved automated web crawling of multiple websites. 
More specifically, we included papers that performed automated web crawling on multiple websites using automated browsers and excluded papers that relied on command line tools or existing datasets without novel data collection. 
We decided to exclude command line tools, as most security and privacy web measurement studies use real browsers, prominently the OpenWPM framework\footnote{OpenWPM, ``Studies,'' \url{https://github.com/openwpm/studies/blob/main/studies.md}}. 
Furthermore, none of our browser-based detection vectors (C8--C12) would apply, diluting our results. 

We evaluated papers on twelve criteria organized in two dimensions: General reproducibility (C1--C6) and bot detection awareness (C7--C12). 
Our framework builds on the reproducibility criteria established by Demir et al.~\cite{DBLP:conf/www/DemirGUWHP22}, who surveyed 117 crawling papers (2016--2021) and identified 18 documentation criteria. 
Notably, their criterion C8 (``State bot detection evasion approach'') exhibited the highest omission rate at 88\%, suggesting that detection-related methodology is systematically under-documented. 
We retained six core reproducibility criteria adapted from Demir et al. and expanded their single bot detection criterion into six sub-criteria that separately assess detection outcomes and mitigation methods. 
This decomposition enables finer-grained analysis of which detection-related aspects researchers document versus omit. We employed a categorical coding scheme summarized in Table~\ref{tab:ratings}.

\begin{table}[!t]
\centering
\caption{Rating scale for criterion evaluation.}
\label{tab:ratings}
\small
\begin{tabular}{@{}l>{\raggedright\arraybackslash}p{9.5cm}@{}}
\toprule
\textbf{Rating} & \textbf{Description} \\
\midrule
Satisfies & The paper provides sufficient detail to replicate the methodology (e.\,g., exact tool names, version numbers, parameter values, or links to artifacts) \\
Partial & The paper acknowledges the aspect but lacks actionable detail (e.\,g., ``used a custom WebDriver'' without implementation details, ``added delays'' without timing values) \\
Omit & The paper does not address this aspect \\
\bottomrule
\end{tabular}
\end{table}

All papers were independently coded by the first author.
To assess coding reliability, a stratified subset of 20 papers (25\%) was independently double-coded by the second author, yielding a Cohen's $\kappa$ of 0.90 (almost perfect agreement) across 260 rating pairs.
We attribute the high agreement to the predominantly objective nature of our criteria (e.\,g., whether a tool is named or a URL is provided), which leaves limited room for subjective interpretation.
The 16 disagreements (6.2\%) were resolved through discussion.

\subsubsection{Reproducibility Criteria (C1--C6)}

Criteria C1--C3 and C6 are self-explanatory from Table~\ref{tab:criteria}.
We clarify two criteria with non-obvious thresholds:

\paragraph{C4: Browser Documented.} A paper satisfies this criterion if browser and version number are named (e.\,g., ``Firefox 140'').
Papers only mentioning the browser or browser family receive partial credit.

\paragraph{C5: Dataset Documented.} A paper satisfies this criterion if it names the site list source (e.\,g., Tranco, Alexa, CrUX), quantifies the scope (N sites or pages), and provides a version or snapshot date for the list (e.\,g., Tranco list ID, Alexa retrieval date).
Papers that name the source and scope but omit the list version receive partial credit.

\subsubsection{Bot Detection Awareness Criteria (C7--C12)} 
These criteria assess transparency regarding bot detection risks and countermeasures. 
They are grounded in known detection vectors documented in evasion frameworks~\cite{DBLP:conf/conext/KrumnowJK22}\footnote{berstend, ``puppeteer-extra-plugin-stealth,'' \url{https://github.com/berstend/puppeteer-extra}; DuckDuckGo, ``tracker-radar-collector,'' \url{https://github.com/duckduckgo/tracker-radar-collector}} and community resources\footnote{AbrahamJuliot, ``CreepJS,'' \url{https://github.com/AbrahamJuliot/creepjs}; Vastel, ``Detecting Selenium Chrome,'' \url{https://datadome.co/bot-management-protection/detecting-selenium-chrome/}} that enumerate automation-specific artifacts, browser fingerprinting surfaces, and behavioral signals.
Criterion C7 captures detection outcomes, while C8--C12 capture mitigation methods. 

\paragraph{C7: Failure Rate Reported.} Whether the paper reports failures or detection incidents. 
This criterion has two components: 

\begin{enumerate}
\item \textit{C7a: Error Rate Reported.} General scan failures (e.\,g., ``X\% of scans failed'', ``N timeouts''). 
Qualitative mentions receive partial credit. 

\item \textit{C7b: Blocking Rate Reported.} Explicit discussion of bot detection or blocking (e.\,g., ``X\% of sites blocked our crawler'', ``N CAPTCHAs encountered''). 
Qualitative mentions receive partial credit. 
We note that omission is ambiguous: it may indicate no failures occurred, that failures were unreported, or that failed sites were silently excluded. 
\end{enumerate}

\paragraph{C8: Display Mode Documented.} Whether the paper specifies headless versus headed browser execution. 
A paper satisfies this criterion by explicitly stating the display mode. 
Implied references (``ran without display'', ``Chrome extension'', or tools that default to headless mode such as Puppeteer) receive partial credit.
Note that we also treat ``crawling with a Chrome extension'' as ambiguous and therefore award only partial credit: 
Historically, Chrome could run extensions only in headed mode, but extension support in headless mode was introduced in 2023.\footnote{\url{https://github.com/puppeteer/puppeteer/issues/659}; \url{https://developer.chrome.com/blog/removing-headless-old-from-chrome}}
Without an explicit statement, or a reproducible configuration description, it is not possible to determine whether the crawl was performed with a visible display or in headless mode. 

\paragraph{C12: Stealth Measures Documented.} Criteria C9--C11 are self-explanatory from Table~\ref{tab:criteria}. For C12, rate limiting, user agent configuration, and display mode are excluded as they are captured separately in C8--C10.

\begin{table}[!t]
\centering
\caption{Criteria for evaluating Web measurement studies.}  
\label{tab:criteria}
\small
\begin{tabular}{@{}ll>{\raggedright\arraybackslash}p{4.0cm}p{6.6cm}@{}}
\toprule
& \textbf{ID} & \textbf{Criterion} & \textbf{Description} \\
\midrule
\multirow{9}{*}{\rotatebox[origin=c]{90}{Reproducibility}}
& C1 & Crawler Available & Crawling implementation is publicly accessible (e.\,g., GitHub, Zenodo) \\
& C2 & Data Available & Collected data is publicly accessible \\
& C3 & Tool Documented & Crawling technology is named (e.\,g., Selenium, Puppeteer, OpenWPM) \\
& C4 & Browser Documented & Browser is specified for browser-based crawlers \\
& C5 & Dataset Documented & Site list source, scope, and version are specified (e.\,g., Tranco ID N7QVW, top 10k) \\
& C6 & Crawl Date Reported & Temporal information is provided \\
\midrule
\multirow{10}{*}{\rotatebox[origin=c]{90}{Bot Detection}}
& C7a & Error Rate Reported & General scan failures are quantified \\
& C7b & Blocking Rate Reported & Bot detection or blocking incidents are discussed \\
& C8 & Display Mode Documented & Headless vs.\ headed execution is specified \\
& C9 & User Agent Documented & HTTP User-Agent configuration is specified \\
& C10 & Rate Limiting \mbox{Documented} & Request timing and throttling are specified \\
& C11 & IP Infrastructure \mbox{Documented} & Network infrastructure is named (e.\,g., AWS, university network) \\
& C12 & Stealth Measures \mbox{Documented} & Active evasion techniques are specified (e.\,g., stealth plugins) \\
\bottomrule
\end{tabular}
\end{table}

\subsection{Results}\label{sec:litsurvey:results}

Having screened the titles and abstracts of 132 papers across seven venues, we identified 81 that match our inclusion criteria (PETS: 21, WWW: 17, IMC: 14, S\&P: 10, CCS: 9, USENIX: 8, NDSS: 2), published between 2020 and 2025.
Table~\ref{tab:survey-results} summarizes the results.

\begin{table}[htbp]
\centering
\caption{Criterion evaluation results across 81 papers.}
\label{tab:survey-results}
\small
\begin{tabular}{@{}ll rr@{\hspace{1.5em}}rr@{\hspace{1.5em}}rr@{}}
\toprule
& \textbf{Criterion} & \multicolumn{2}{c@{\hspace{1.5em}}}{\textbf{Satisfies}} & \multicolumn{2}{c@{\hspace{1.5em}}}{\textbf{Partial}} & \multicolumn{2}{c}{\textbf{Omit}} \\
& & $n$ & \% & $n$ & \% & $n$ & \% \\
\midrule
\multirow{6}{*}{\rotatebox[origin=c]{90}{Reprod.}}
& C1: Crawler Available    & 41 & 51 &  4 &  5 & 36 & 44 \\
& C2: Data Available       & 23 & 28 &  7 &  9 & 51 & 63 \\
& C3: Tool Documented      & 80 & 99 &  1 &  1 &  0 &  0 \\
& C4: Browser Documented   & 37 & 46 & 42 & 52 &  2 &  2 \\
& C5: Dataset Documented   & 47 & 58 & 34 & 42 &  0 &  0 \\
& C6: Crawl Date Reported  & 42 & 52 & 11 & 14 & 28 & 35 \\
\midrule
\multirow{7}{*}{\rotatebox[origin=c]{90}{Bot Det.}}
& C7a: Error Rate Reported      & 37 & 46 &  9 & 11 & 35 & 43 \\
& C7b: Blocking Rate Reported   &  4 &  5 & 10 & 12 & 67 & 83 \\
& C8: Display Mode Documented   & 25 & 31 & 18 & 22 & 38 & 47 \\
& C9: User Agent Documented     & 11 & 14 &  5 &  6 & 65 & 80 \\
& C10: Rate Limiting Documented &  6 &  7 &  2 &  2 & 73 & 90 \\
& C11: IP Infrastructure Doc.   & 29 & 36 & 16 & 20 & 36 & 44 \\
& C12: Stealth Measures Doc.    & 19 & 23 &  7 &  9 & 55 & 68 \\
\bottomrule
\end{tabular}
\end{table}

\subsubsection{Reproducibility (C1--C6)}
Tool choice (C3) is nearly universally documented (99\%), with OpenWPM being the most common framework (25 papers), followed by Selenium (16), Puppeteer (14), and custom implementations (12).
Dataset documentation (C5) reaches 58\% when requiring a versioned or dated site list, with Tranco (40 papers) having largely replaced Alexa (24) as the preferred toplist.
Browser choice (C4) is named in 98\% of papers, though only 46\% specify the version number.

Code and data availability remain limited: while a slight majority of papers release their crawling code (C1: 51\%), only 28\% share their collected data (C2).
Crawl dates (C6) are reported by 52\%, with another 14\% providing only vague temporal references.

\subsubsection{Bot Detection Awareness (C7--C12)}
The bot detection criteria reveal a systematic documentation gap.
The most striking finding is C7b: only 5\% of papers explicitly quantify bot detection or blocking rates, and 83\% omit any discussion of blocking entirely.
This is notable because, as our measurement study demonstrates (Section~\ref{sec:measurement}), bot detection is deployed on a substantial fraction of popular websites and can silently affect crawl results.

General error rates (C7a) fare better, with 46\% providing quantified failure counts, though 43\% still omit error reporting entirely.
Display mode (C8) is omitted by 47\% of papers, despite headless execution being a primary bot detection signal.
User-Agent configuration (C9) is documented in only 20\% of papers.
Rate limiting (C10) has the highest omission rate at 90\%.
IP infrastructure (C11) is documented in 56\% of papers, often because studies use cloud vantage points as part of their methodology.
Active stealth measures (C12) are documented in 32\% of papers, with the remainder neither employing nor discussing evasion techniques.

Across the seven bot detection criteria (C7a--C12), papers address an average of only 2.4 out of 7 criteria.
Eight papers (10\%) do not address any bot detection criterion at all.
Only a single paper addresses six criteria; none addresses all seven.

%% file: sections/dns-census.tex
\section{Prevalence of Content Distribution Networks}\label{sec:dns-census}

To quantify the potential impact of bot detection on web measurements, we performed DNS resolution on the Tranco Top 1 Million domains and mapped them to infrastructure providers.
Major CDNs such as Cloudflare and Akamai also offer bot detection services, making CDN prevalence a proxy for bot detection exposure.

\subsection{Data Collection}

As a data source, we used the Tranco Top 1 Million list~\cite{DBLP:conf/ndss/PochatGTKJ19} from February 25, 2026 (ID: GVL9K), as it is one of the most widely used top lists in web measurement research that we encountered during our literature survey.
We resolved all A records for all domains using a parallelized Python program with dnspython, configured with three public DNS resolvers from Google (8.8.8.8), Cloudflare (1.1.1.1), and Quad9 (9.9.9.9) for redundancy. 

Of the 1 million domains, 893{,}411 (89.3\%) resolved to at least one A record, yielding 1{,}472{,}477 A records in total.
The remaining 106{,}589 domains returned NXDOMAIN or timed out, consistent with parked, expired, or misconfigured entries commonly found in aggregated toplists~\cite{DBLP:conf/ndss/PochatGTKJ19}.

\subsection{Analysis Method}

We mapped the resolved IP addresses to their owning autonomous system (AS) using the IPinfo Lite dataset (CSV snapshot 2026-01-29). 
Of 1{,}472{,}477 A records, 1{,}380{,}865 (93.8\%) were successfully matched to an AS owner.
Where multiple AS names refer to the same organization (e.\,g., ``Cloudflare London, LLC'' and ``Cloudflare, Inc.''), we consolidated them into a single entity. 
Note that we count A records rather than unique domains, so our percentages reflect infrastructure footprint: providers whose domains resolve to multiple A records (e.\,g., for load balancing) are weighted proportionally higher.

\subsection{Results}

\begin{figure}
\includegraphics[width=\textwidth]{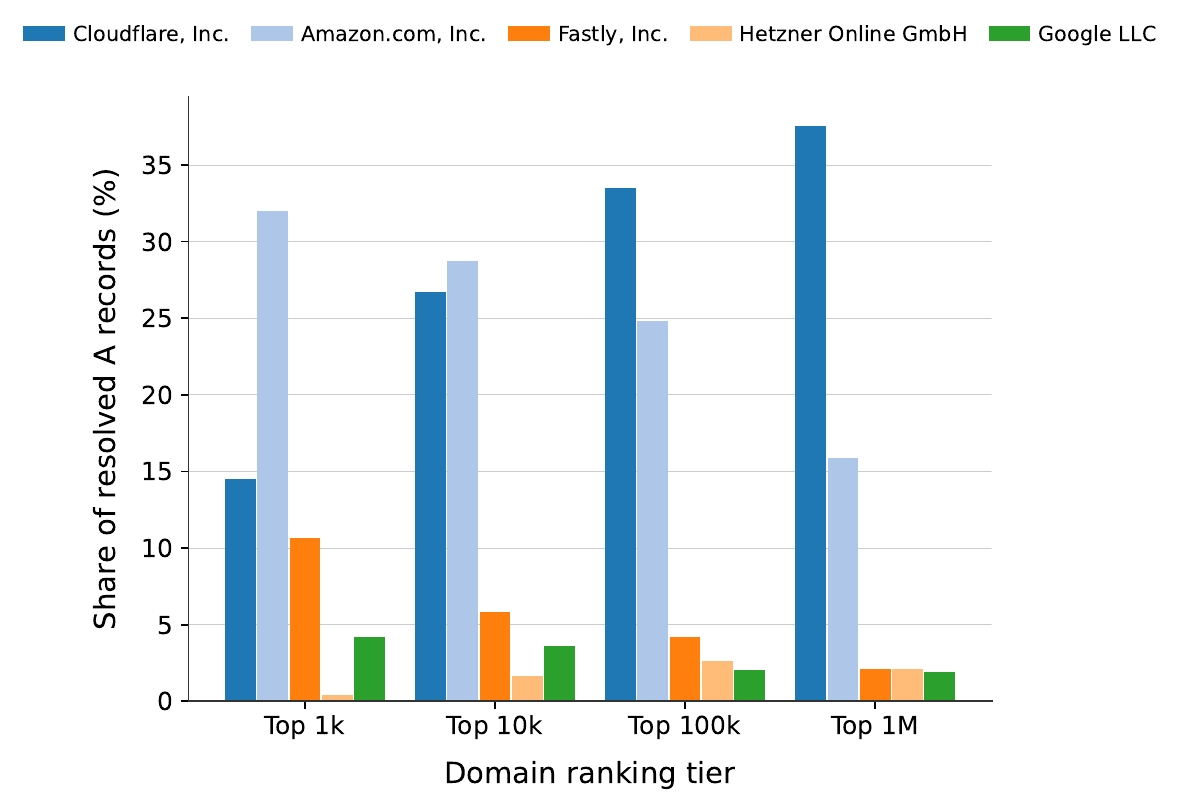}
\caption{AS ownership share across Tranco domain ranking tiers (top 5 providers).}\label{fig:as-ownership}
\end{figure}

Our analysis reveals a significant concentration of web infrastructure among a small number of providers, which is in line with previous work~\cite{DBLP:conf/imc/KashafSA20}. 
Cloudflare alone accounts for 40.0\% of AS-matched A records, followed by Amazon at 16.9\%.
The top 3 providers cover 59.2\% and the top 5 cover 63.5\%.
Note that Figure~\ref{fig:as-ownership} uses all resolved A records as the denominator (including the 6.2\% not matched to an AS), yielding slightly lower percentages.

As shown in Figure~\ref{fig:as-ownership}, this concentration shifts across popularity tiers.
Amazon dominates among the top 1k at 32.0\% of resolved A records, but its share decreases to 15.9\% across the full million.
Cloudflare shows the inverse trend, growing from 14.5\% in the top 1k to 37.6\% in the top 1M.
This suggests that the most popular websites rely more heavily on enterprise-grade CDN and cloud providers, while lower-ranked sites increasingly adopt Cloudflare, likely due to its accessible free and low-cost tiers which include bot detection by default.

Table~\ref{tab:as_domain_count} combines infrastructure prevalence with blocking outcomes from our measurement study (Section~\ref{sec:measurement}).
Several of these providers offer bot detection services, notably Cloudflare (Bot Management, enabled by default on free plans), Akamai (Bot Manager), and Fastly (Signal Sciences); however, hosting on a capable provider does not guarantee active detection, as services like AWS WAF are opt-in and paid.
The blocking rates confirm that infrastructure provider is the primary correlate of bot detection: Cloudflare (37.0\%) and Akamai (26.4\%) block at rates far above the overall average, while providers without default-on bot detection range from 0--16\%.
Notably, Akamai's high block rate is almost entirely unclassified by vendor signatures in the HTTP responses (103 of 110 blocked sites), consistent with Akamai serving generic error pages without vendor branding. 
This means that researchers receiving these responses have no straightforward way to distinguish bot detection blocks from legitimate server errors, making silent sample loss particularly difficult to detect.

\begin{table}[htbp]
\centering
\caption{Top 10 Autonomous Systems by A record count (Tranco Top~1M) with blocking rates from our Top~10K scan. 
Block rate: percentage of sites in the Top~10K hosted on this AS that received HTTP 403/429/503 in at least one browser condition (prevalence is computed on Tranco Top 1M, blocking rates on the Top 10K sample). 
Rates shown only for providers with ${\geq}100$ sites in the Top~10K.}
\label{tab:as_domain_count}
\begin{tabular}{@{}lrr@{\hspace{2em}}rrr@{}}
\toprule
\textbf{AS Name} & \textbf{A records} & \textbf{\%} & \textbf{Sites (10K)} & \textbf{Blocked} & \textbf{Block rate} \\
\midrule
Cloudflare, Inc.            & 553{,}014 & 40.0 & 2{,}128 & 788 & 37.0 \\
Amazon.com, Inc.            & 233{,}600 & 16.9 & 1{,}720 & 280 & 16.3 \\
Fastly, Inc.                &  31{,}470 &  2.3 &    321  &  50 & 15.6 \\
Hetzner Online GmbH         &  30{,}746 &  2.2 &     91  &  4 &   --- \\
Google LLC                  &  28{,}140 &  2.0 &    531  &  27 &  5.1 \\
Akamai Connected Cloud      &  23{,}802 &  1.7 &    417  & 110 & 26.4 \\
OVH SAS                     &  21{,}825 &  1.6 &     81  &  3 &   --- \\
Microsoft Corporation       &  18{,}874 &  1.4 &    162  &  24 & 14.8 \\
Hostinger International Ltd &  13{,}799 &  1.0 &      1  &  0 &   --- \\
DigitalOcean, LLC           &  11{,}071 &  0.8 &     22  &  4 &   --- \\
\bottomrule
\end{tabular}
\end{table}

%% file: sections/bot-detection-measurement.tex
\section{Prevalence of Bot Detection and Impact on Web Measurement Studies}\label{sec:measurement}

Having established that bot detection reporting is systematically absent from web measurement research (Section~\ref{sec:literature-survey}) and that a majority of popular websites are served through CDN infrastructure with bot detection capabilities (Section~\ref{sec:dns-census}), we now address our third contribution.
We quantify how often bot detection actually blocks automated browsers and characterize the detection techniques and services involved.

\subsection{Methods}\label{sec:methods}

\subsubsection{Instrumentation Approach.} While previous works instrumented individual APIs, we take a different approach. 
Before page scripts execute, we inject JavaScript that dynamically configures proxies for \emph{all} property accesses and function calls. 
A proxy is a JavaScript metaprogramming feature that allows us to intercept and log all property accesses on an object. 
This maximizes API coverage without relying on a predefined list of instrumentation targets. 
We additionally flag APIs associated with browser fingerprinting and automation detection, derived from the following sources: 

\begin{enumerate}
\item \textbf{WPM\textsubscript{hide}}~\cite{DBLP:conf/conext/KrumnowJK22}, an evasion module developed for OpenWPM
\item \textbf{puppeteer-extra-plugin-stealth}\footnote{berstend, ``puppeteer-extra-plugin-stealth,'' \url{https://github.com/berstend/puppeteer-extra/tree/master/packages/puppeteer-extra-plugin-stealth}}, an evasion library that has been used in past research~\cite{DBLP:conf/ccs/BuiTS22}
\item \textbf{Tracker Radar Collector}\footnote{DuckDuckGo, ``tracker-radar-collector'', \url{https://github.com/duckduckgo/tracker-radar-collector}}, a puppeteer-based crawler featuring evasion methods that collects third party request data, also used in previous research~\cite{DBLP:conf/sp/MotiSBBMMA24}
\item \textbf{Documented Detection Techniques}: Properties identified in research~\cite{DBLP:conf/imc/JueckstockK19} and community resources
\footnote{Abraham Juliot, ``CreepJS'', \url{https://github.com/abrahamjuliot/creepjs}; \url{https://stackoverflow.com/a/41220267}}
\end{enumerate}

From these four sources, we compiled 132 JavaScript properties. 
Accesses to one or more of these properties signal that a site may be attempting to detect automated browsers. 
We categorize the properties into three confidence tiers based on their specificity to automation detection, as many also serve legitimate fingerprinting purposes:

\begin{enumerate}
\item \textbf{Tier 1}: No legitimate use case except for automation detection. 
These are browser properties that are injected by automation frameworks (e.\,g., \texttt{navigator.\allowbreak{}webdriver}, \texttt{window.\allowbreak{}\_selenium}).
\item \textbf{Tier 2}: Strong automation detection indicators that also serve other use cases. For instance, \texttt{chrome.\allowbreak{}runtime} and \texttt{navigator.\allowbreak{}plugins} can be used to fingerprint a browser or to detect discrepancies in its fingerprint, which would indicate an automated browser.
\item \textbf{Tier 3}: APIs that are generally used for fingerprinting browsers, but could also be used to detect browser automation (e.\,g., \texttt{navigator.\allowbreak{}userAgent}).
\end{enumerate}

Tier 1 accesses represent unambiguous detection attempts, while Tier 3 accesses could also stem from regular fingerprinting and must be interpreted in context.

\subsubsection{Instrumentation Architecture.}\label{sec:instrumentation}

Our instrumentation operates at three levels:

\begin{itemize}
\item \textbf{Property interception.} We install a recursive \texttt{Proxy} on the \texttt{window} object whose \texttt{get} trap intercepts property accesses, logs them, and returns nested proxies that track the full property path (e.\,g., \texttt{navigator.\allowbreak{}permissions.\allowbreak{}query}).
\item \textbf{Prototype patching.} For non-configurable properties on built-in prototypes (e.\,g., \texttt{Document.\allowbreak{}prototype}, \texttt{Navigator.\allowbreak{}prototype}, \texttt{RTC\allowbreak{}Peer\allowbreak{}Connection.\allowbreak{}prototype}), we replace prototype getters and methods with instrumented versions that log accesses before delegating to the original implementation.
\item \textbf{Honeypot properties.} Bot detection scripts often check for properties that automation frameworks inject (e.\,g., \texttt{window.\allowbreak{}\_selenium}). These properties do not exist in our browser, but by planting them as invisible traps that log any access attempt and return \texttt{undefined}, we can detect when a site is probing for automation without revealing our browser as automated.
\end{itemize}

While signal tiers and instrumentation methods are related, they are not equivalent.
Tiers reflect semantic confidence (what an API access implies about detection intent), whereas instrumentation methods reflect technical constraints (how we capture the access).
For instance, Tier 1 includes both honeypot properties that do not exist in real browsers (e.\,g., \texttt{window.\allowbreak{}\_selenium}) and real automation artifacts that require property interception (e.\,g., \texttt{navigator.\allowbreak{}webdriver}).
Each logged access records: the full property path (symbol), operation type (\texttt{GET}/\texttt{HAS}/\texttt{SET}/\texttt{CALL}), caller script location, timestamp, and assigned confidence tier. 
When an API is accessed more than 500 times, we record a \texttt{LIMIT\_REACHED} event, a threshold used in prior work to detect Canvas and performance-based fingerprinting~\cite{DBLP:conf/sp/IqbalES21}\footnote{\url{https://github.com/UmarIqbal/OpenWPM/blob/9e6234ef7a542c5d4bf56493eb3b39840c7dfecb/automation/Extension/firefox/data/content.js}}. 
Additionally, all scripts are stored with their URL and script content for \emph{bot detection vendor} attribution. 

\subsubsection{Experimental Conditions.}

We examine two factors: browser engine (Firefox and Chromium) and display mode (headless and headed). 
Previous work and industry practice suggest that headless browsers exhibit detectable properties, and that browser engines differ in their exposed API sets. 
For instance, Firefox removed the Battery API in version~52 to reduce fingerprinting by trackers~\cite{DBLP:conf/esorics/OlejnikACD15}, and such browser engine differences can themselves be used as fingerprinting signals~\cite{DBLP:conf/sp/IqbalES21}. 
This $2 \times 2$ design yields four conditions, each applied to every site in our sample. 
All conditions use Playwright for consistent automation semantics across browser engines.
While Playwright itself is detectable, our instrumentation executes before page scripts and all four conditions share the same automation framework, so any Playwright-specific bias affects all conditions equally.
As we discuss below, scanning from a datacenter IP rather than a residential network affects absolute blocking rates, but our paired design still allows us to observe meaningful differences between the $2 \times 2$ conditions.
To minimize confounds, all scans are run from a single Hetzner datacenter in Nuremberg, ensuring a uniform region and ASN type, with IP rotation between batches and randomized scan order to avoid repeated same-site requests from the same IP in quick succession.
Table~\ref{tab:crawler-config} summarizes our crawler configuration.

\begin{table}[t]
\centering
\caption{Crawler configuration summary.}
\label{tab:crawler-config}
\small
\begin{tabular}{@{}ll@{}}
\toprule
\textbf{Parameter} & \textbf{Value} \\
\midrule
Automation framework  & Playwright 1.57.0 \\
Browser engines       & Chromium 143.0.7499.4, Firefox 144.0.2 \\
Display modes         & Headless, headed \\
Site list             & Tranco Top~10K (ID: GVL9K, 2026-02-25) \\
Scan date             & 2026-02-27 to 2026-03-02 (spoofing: 2026-03-03) \\
User-Agent            & Playwright defaults (unmodified) \\
Rate limiting         & IP rotation between batches of 1{,}000 sites \\
IP infrastructure     & Hetzner Cloud, Nuremberg, DE (datacenter) \\
Stealth measures      & None (default Playwright configuration) \\
\bottomrule
\end{tabular}
\end{table}

\subsubsection{Site Selection and Sampling.}

We draw the first 10{,}000 sites of the Tranco Top 1 Million list~\cite{DBLP:conf/ndss/PochatGTKJ19}. 
As in Section~\ref{sec:dns-census}, we use the Toplist from February 25, 2026. 
Sites are shuffled deterministically with a documented seed value for reproducibility purposes and partitioned into 10 batches of 1{,}000 sites each.
Within each batch, we generate 4{,}000 scans (1{,}000 sites $\times$ 4 conditions), again shuffled with a documented per-batch seed to distribute same-site scans temporally. 

\subsubsection{Execution Protocol.}

Scans proceed in batches with IP rotation between them. 
After each batch, we delete the virtual machine along with its public IP address and restore it from a snapshot, thus receiving a different IP address. 
Each scan is executed as follows:

\begin{enumerate}
\item \textbf{Navigation}: Load the page with a 30-second timeout
\item \textbf{Idle Detection}: Wait until no new API accesses have been recorded for 2 seconds, or 30 seconds have passed in total. Meanwhile, all properties are logged as described previously
\item \textbf{Terminate Scan}: Finally the browser saves a screenshot to the file system and marks the scan as successful. Screenshots serve as ground truth for manual verification of blocking responses (Section~\ref{sec:rq1})
\end{enumerate}

Our study minimizes harm and impact by only visiting publicly accessible homepages and not performing interactions beyond the front page load. We do not collect personal data.

This design ensures all four conditions for a given site execute under the same IP address and within a narrow time window, enabling paired comparisons that control for IP reputation and site content variability. 
We specifically rotate IPs to mitigate known measurement artifacts, as previous research demonstrates that web scanners frequently encounter ``refusenik'' websites that systematically drop connections or alter content based on the degraded reputation of the scanner's network vantage point~\cite{DBLP:conf/dimva/AzadSLN20,DBLP:conf/IEEEares/GundelachH23,DBLP:conf/www/JueckstockSSBPV21}. 

\paragraph{Methodological caveats.}
Three aspects of our instrumentation affect interpretation.
First, our JavaScript proxies create detectable artifacts; sophisticated detection scripts can identify proxied objects via \texttt{Function.\allowbreak{}prototype.\allowbreak{}toString()} inspection or timing side-channels.
However, this primarily affects our API access counts (Section~\ref{sec:rq2}), not our blocking measurements: HTTP-level blocking decisions (403/429/503) are made by CDN infrastructure independently of client-side JavaScript.
The observed headless-vs-headed blocking asymmetry (Section~\ref{sec:rq1}) and header spoofing results (Section~\ref{sec:rq1}) are based on HTTP-level outcomes, not API counts, and are therefore not affected by instrumentation detectability.
Second, we capture only client-side JavaScript detection; server-side mechanisms (IP reputation, TLS fingerprinting, behavioral analysis) are outside our observational scope, making our prevalence estimates a lower bound.
Third, observing an API access does not establish that a site acted on the result: a site may probe \texttt{navigator.\allowbreak{}webdriver} for telemetry without blocking.
We correlate signal access with scan outcomes but cannot establish definitive causal links; only Tier~1 signals are unambiguously detection-related.
We discuss further limitations in Section~\ref{sec:discussion}.

\subsection{Results}

We present our findings across two research questions: detection prevalence and cross-condition differences (RQ1), and the detection landscape including techniques and services (RQ2).
Because not every site completed all four conditions (due to DNS failures, timeouts, or SSL errors), denominators vary by analysis.
RQ1 reports per-condition totals out of 10{,}000~sites each, with paired comparisons on the 6{,}936 Chromium and 7{,}063 Firefox sites where both headless and headed conditions returned an HTTP response.
API prevalence in RQ2 uses the 7{,}944~sites with at least one successful scan that produced property-access data.

\subsection{Detection Prevalence (RQ1)}\label{sec:rq1}

Table~\ref{tab:detection-classification} shows the detection classification across all four browser conditions.
We classify each scan based on its HTTP status code and scan outcome into \emph{success} (any 2xx response), \emph{soft block} (HTTP 403 Forbidden, 429 Too Many Requests, or 503 Service Unavailable), and \emph{timeout} (page did not respond within 30\,s).
Not all soft blocks necessarily represent bot detection, so we validate this classification below by analyzing response bodies and cross-condition blocking patterns.

\begin{table}[t]
\centering
\caption{Detection classification by browser condition ($n{=}10{,}000$ sites per condition).
Not shown: ${\sim}2{,}100$--$2{,}450$ infrastructure failures (DNS, SSL, connection errors) per condition, screenshot timeouts (${\sim}440$ Chromium, ${\sim}150$ Firefox), and ${\sim}170$--$220$ other HTTP status codes (3xx, 500, etc.).}
\label{tab:detection-classification}
\begin{tabular}{@{}l rr@{\hspace{1.5em}}rr@{\hspace{1.5em}}rr@{}}
\toprule
\textbf{Condition} & \multicolumn{2}{c@{\hspace{1.5em}}}{\textbf{Success (2xx)}} & \multicolumn{2}{c@{\hspace{1.5em}}}{\textbf{Soft block}} & \multicolumn{2}{c}{\textbf{Nav.\ timeout}} \\
& $n$ & \% & $n$ & \% & $n$ & \% \\
\midrule
Chromium headless & 5{,}550 & 55.5 & 1{,}521 & 15.2 & 117 & 1.2 \\
Chromium headed   & 6{,}435 & 64.4 &    721 &  7.2 & 109 & 1.1 \\
Firefox headless  & 6{,}462 & 64.6 &    678 &  6.8 &  92 & 0.9 \\
Firefox headed    & 6{,}457 & 64.6 &    677 &  6.8 &  98 & 1.0 \\
\bottomrule
\end{tabular}
\end{table}

Chromium headless is a clear outlier at 15.2\%, while the remaining three conditions cluster between 6.8--7.2\%.
This gap is driven by the interaction of browser engine and display mode, not by either factor alone: among the 6{,}936 Chromium sites where both headless and headed conditions completed, 772~sites are blocked exclusively in headless mode compared to only 14~blocked exclusively in headed mode. Among the 7{,}063 paired Firefox sites, the headless/headed difference is negligible (15 vs.\ 12 sites).

A key contributing factor is the \texttt{sec-ch-ua} Client Hints header, which exposes \texttt{HeadlessChrome} as a brand value in Chromium's headless mode. Firefox does not implement Client Hints, eliminating this detection vector entirely. Additionally, Playwright's Chromium headless User-Agent string includes ``HeadlessChrome'' and exposes the full build number (\texttt{143.0.7499.4}) rather than the reduced form used in headed mode (\texttt{143.0.0.0}).

\paragraph{Header spoofing experiment.}
To establish causality, we re-crawled the 798~sites that received HTTP~403 exclusively in Chromium headless with spoofed HTTP headers: we replaced \texttt{HeadlessChrome} with \texttt{Chrome} in the User-Agent string and removed the \texttt{HeadlessChrome} brand from the \texttt{sec-ch-ua} header, matching headed Chromium's headers exactly.
Of 784~sites that returned a response, 590~(75\%) returned HTTP~2xx, while 194~(25\%) remained blocked, indicating deeper detection layers (e.\,g., JavaScript fingerprinting, behavioral analysis, or TLS fingerprinting).
Consistent with this, our instrumentation shows that \texttt{navigator.\allowbreak{}user\allowbreak{}Agent\allowbreak{}Data} (the Client Hints JavaScript API) is probed on 730 of the originally blocked sites, and \texttt{navigator.userAgent} on 489, suggesting that both detection signals are widely deployed.

\paragraph{Same infrastructure, escalated response for headless.}
Of the 1{,}554~sites soft-blocked in at least one condition, 1{,}521~(97.9\%) are blocked in \emph{Chromium headless}, while the three other conditions each see 677--721 blocks.
Of the 750~sites blocked in at least one non-\emph{Chromium headless} condition, 717~(95.6\%) are also blocked in \emph{Chromium headless}, confirming that the same infrastructure drives blocking across conditions but escalates its response for headless Chromium.

\paragraph{82\% of blocks are attributable to bot detection.}
To validate that soft blocks represent active bot detection, we classify response bodies by matching their content against known patterns of bot detection vendors (e.\,g., Cloudflare challenge pages, DataDome block pages).
Table~\ref{tab:block-classification} shows the results by unique site.
Of the 1{,}554 sites with at least one soft block, we identify a known vendor on 919~(59.1\%).
The remaining 635 sites lack vendor signatures. To distinguish bot detection from infrastructure-level errors, we examine cross-condition behavior: 282~sites (18.1\%) are blocked uniformly across every condition in which they responded, potentially resulting from geographic restrictions, ASN-based blocks, or server misconfigurations.
The remaining 353~sites (22.7\%) show condition-dependent blocking, with 334~(94.6\%) blocked in Chromium headless, consistent with the header-level detection mechanism established above.
We manually verified all vendor-confirmed classifications by inspecting their screenshots and response bodies.
Combining vendor-confirmed and condition-dependent blocks, we attribute 1{,}272~sites (81.9\%) to bot detection, with 282~(18.1\%) representing infrastructure-level access denial.

\begin{table}[ht]
\centering
\caption{Bot detection vendor signatures in HTTP 403/429/503 response bodies, by unique site. Each site is assigned to the highest-priority vendor match across all four conditions.}
\label{tab:block-classification}
\begin{tabular}{@{}lrr@{}}
\toprule
\textbf{Category} & \textbf{Sites} & \textbf{\%} \\
\midrule
Cloudflare (challenge) & 792 & 51.0 \\
DataDome               &  65 &  4.2 \\
PerimeterX/HUMAN       &  26 &  1.7 \\
CAPTCHA                &  23 &  1.5 \\
Incapsula/Imperva      &  13 &  0.8 \\
\cmidrule{2-3}
\textbf{Subtotal (vendor-confirmed)} & \textbf{919} & \textbf{59.1} \\
\midrule
Unclassified (condition-dependent) & 353 & 22.7 \\
Unclassified (uniform)             & 282 & 18.1 \\
\midrule
\textbf{Total}                     & \textbf{1{,}554} & \textbf{100.0} \\
\bottomrule
\end{tabular}
\end{table}

\paragraph{Blocking is provider-driven, not rank-driven.}
Blocking does not correlate with site popularity (Mann-Whitney~$U$, $p < 0.001$, $r = -0.07$) but with infrastructure provider: as shown in Table~\ref{tab:as_domain_count}, Cloudflare (37\%) and Akamai (26\%) block at the highest rates, while providers without default-on bot detection range from 0--16\%.

\subsection{Detection Landscape (RQ2)}\label{sec:rq2}

While 75\% of current blocks rely on header-level signals, the JavaScript-based detection infrastructure already deployed is more extensive than current blocking rates reflect.
Our proxy-based instrumentation (Section~\ref{sec:instrumentation}) captures not only which APIs are probed, but also which scripts perform the probing, enabling caller-level attribution of detection activity.
The most common Tier~3 APIs (e.\,g., \texttt{navigator.userAgent}, \texttt{localStorage}) are each accessed on over 75\% of sites but reflect standard web development practice.
We focus first on Tier~1 signals, which have no legitimate use case except automation detection, then examine how vendors combine signals across tiers into integrated detection pipelines.

\paragraph{34\% of sites probe for automation artifacts.}
The most prevalent Tier~1 signal is \texttt{navigator.\allowbreak{}webdriver}, accessed on 2{,}730~(34\%) of 7{,}944 successfully scanned sites.
This property returns \texttt{true} in all Playwright-controlled browsers regardless of display mode, making it the single most common bot detection check in our dataset.
We group the 28 individual Tier~1 signals into eight automation framework families, reporting the percentage of 7{,}944~sites that probe at least one signal in each family: WebDriver API~(43\%), PhantomJS~(32\%), NightmareJS~(31\%), WebDriver Protocol~(29\%), Selenium~(28\%), Chrome DevTools~(21\%), Awesomium~(20\%), and ChromeDriver~(8\%).
Across all eight families, 3{,}649~sites (46\%) probe at least one Tier~1 signal, indicating that automation detection is deployed on nearly half of the successfully scanned sites.
Notably, sites still probe for discontinued frameworks such as PhantomJS and NightmareJS, suggesting that detection scripts accumulate checks without removing obsolete ones.

\paragraph{Bot detection scripts rank among the most common third-party scripts.}
Cross-referencing our response body classification (Table~\ref{tab:block-classification}) with script URLs reveals two distinct delivery models.
\emph{Visible} vendors serve detection scripts from identifiable domains: Cloudflare from \texttt{challenges.\allowbreak{}cloudflare.\allowbreak{}com} (705~sites), DataDome from \texttt{ct.\allowbreak{}captcha-\allowbreak{}delivery.\allowbreak{}com} (61~sites), and PerimeterX/HUMAN from \texttt{client.\allowbreak{}px-cloud.\allowbreak{}net} (25~sites).
In contrast, \emph{invisible} vendors such as Incapsula/Imperva proxy detection logic through first-party domains or operate server-side, so script domain analysis alone undercounts detection deployment.
We can therefore only provide a lower bound of bot detection script prevalence by counting scripts from visible vendors.
To do so, we group minified script variants by their lexical structure\footnote{We tokenize scripts with esprima (\url{https://pypi.org/project/esprima/}) and hash the token sequences.}.
Cloudflare Turnstile ranks as the 6th most deployed third-party script on the Tranco Top~10K (794~sites), ahead of major analytics libraries, with Google reCAPTCHA at rank~17 (398~sites).
Bot detection scripts are thus among the most common third-party scripts on the web.

\paragraph{Vendors combine honeypot probing with browser fingerprinting.}
Cross-referencing script callers with both signal tiers and blocking outcomes reveals that major bot detection vendors deploy integrated detection pipelines that combine window-level honeypot probing with multi-category browser fingerprinting within a single script execution (Table~\ref{tab:vendor-pipelines}).
We restrict this analysis to sites that block even headed browsers, representing the most aggressive detection, and report only scripts that probe at least three window-level honeypots.
Note that site counts here reflect script-level attribution and may exceed the response-body counts in Table~\ref{tab:block-classification}, where sites are assigned to a single highest-priority vendor.
Cloudflare's detection scripts probe an average of 13~honeypots and cover all eight fingerprinting categories we track (Canvas, Audio, WebGL, Speech, Fonts, Dimensions, Permissions, Plugins) across 303~headed-blocked sites where our scanner recorded property-access data.
Prior work reverse-engineered a single commercial detector and found it combines fingerprinting with detection~\cite{DBLP:conf/esorics/JonkerKV19}. Our instrumentation confirms this pattern at scale: across hundreds of sites, all three vendors combine honeypot probing and multi-category fingerprinting within a single script execution.

\begin{table}[ht]
\centering
\caption{Bot detection vendor pipelines on headed-blocked sites. Honeypots: number of automation artifacts probed. FP categories: fingerprinting categories exercised (of 8).}
\label{tab:vendor-pipelines}
\begin{tabular}{@{}lrrr@{}}
\toprule
\textbf{Vendor} & \textbf{Blocked sites} & \textbf{Honeypots} & \textbf{FP categories} \\
\midrule
Cloudflare       & 303 & 13 & 8/8 \\
DataDome         &  41 &  4 & 6/8 \\
PerimeterX/HUMAN &  13 &  7 & 8/8 \\
\bottomrule
\end{tabular}
\end{table}

\paragraph{Summary.}
Bot detection blocking is widespread, affecting 7--15\% of the Tranco Top~10K depending on browser configuration, with over 80\% attributable to bot detection rather than infrastructure errors.
The blocking gap is driven by header-level signals that identify Chromium headless, as confirmed by a spoofing experiment that unblocks 75\% of affected sites.
The detection landscape is dominated by a small number of vendors, most prominently Cloudflare, whose scripts combine automation probing with comprehensive browser fingerprinting.